\documentclass[10pt,twocolumn]{article} 

\usepackage[T1]{fontenc}
\usepackage[utf8]{inputenc}

\usepackage{simpleConference}
\usepackage{times}
\usepackage{graphicx}
\usepackage{amssymb}
\usepackage{amsmath}
\usepackage{color,xcolor}
\usepackage{listings}
\usepackage{musicography}
\usepackage{stfloats}
\usepackage{tabularx}
\usepackage{ragged2e}
\usepackage{balance}
\usepackage{cite}
\usepackage[font=small,labelfont=bf,textfont=normal]{caption}

\def\authorname{M. Kanani, S. O'Leary, and J. McDermott}

\usepackage[
  bookmarks=false,
  pdfauthor={\authorname},
  pdfsubject={The Radif Corpus},
  hidelinks
]{hyperref}

\begin{document}

\title{Radif Corpus: A Symbolic Dataset for Non-metric Iranian Classical Music}

\author{
    Maziar Kanani$^1$, Seán O'Leary$^2$, James McDermott$^1$ \\
    $^1$University of Galway, Ireland \\
    $^2$Technological University Dublin, Ireland \\
    \texttt{m.kanani1@universityofgalway.ie} \\
    \texttt{sean.oleary@tudublin.ie} \\
    \texttt{james.mcdermott@universityofgalway.ie}
}

\sloppy 

\maketitle


\begin{abstract}

Non-metric music forms the core of the repertoire in Iranian classical music. \emph{Dastgāhi} music serves as the underlying theoretical system for both Iranian art music and certain folk traditions. At the heart of Iranian classical music lies the \emph{radif}, a foundational repertoire that organizes melodic material central to performance and pedagogy.

In this study, we introduce a digital corpus representing the complete non-metrical \emph{radif} repertoire, covering all 13 existing components of this repertoire. We provide MIDI files (about 281 minutes in total) and data spreadsheets describing notes, note durations, intervals, and hierarchical structures for 228 pieces of music. We faithfully represent the tonality including quarter-tones, and the non-metric aspect. Furthermore, we provide supporting basic statistics, and measures of complexity and similarity over the corpus.

Our corpus provides a platform for computational studies of Iranian classical music. Researchers might employ it in studying melodic patterns, investigating improvisational styles, or for other tasks in music information retrieval, music theory, and computational (ethno)musicology.

\end{abstract}

\section{Introduction}\label{sec:introduction}

While ethnic music traditions from around the world have recently gained more attention in computational research, many still lack the necessary datasets to support such studies. Iran, with its rich diversity of ethnic and folk musical traditions, offers great potential for computational analysis that reflects its regional musical identity.

In this work, we take a step toward by introducing a dataset specifically focused on Iranian non-metric classical music, aiming to support and inspire future studies in this area. We begin by introducing Iranian classical music and its core repertoire, the \emph{radif}. After reviewing previously published datasets, we present our own dataset in detail. Finally, we provide a statistical and visual overview of the dataset, which can serve as a useful reference for researchers and practitioners.

\subsection{Foundations of Iranian Classical Music}

Iranian classical music comes from a larger style of music called \emph{dastgāhi} music. This term describes the theoretical framework underlying Iranian classical music and certain styles of Iranian folk music, such as \emph{bakhtiāri}. The core repertoire of Iranian classical music is \emph{radif} (literally "order"), a structured collection of melodies transmitted across generations and foundational to performance and pedagogy.

\emph{Radif} is a collection of melodies organized into a specific sequence, typically divided into 12 subcategories (traditionally 13). Out of these, seven are primary subcategories known as \emph{dastgāh}, and five (respectively six) are secondary, referred to as \emph{āvāz}, which can also be considered as smaller \emph{dastgāh} and serve as subcategories for the primary seven. Each of these subcategories is known for its distinctive characteristics. They are typically recognized based on their main mode (Introduced in the first \emph{gūsheh}), the functional roles of their tones within that mode, and the specific sequence of \emph{gūshehs} within them.

The \emph{dastgāhs} are: \emph{shur}, \emph{segāh}, \emph{navā}, \emph{homāyūn}, \emph{chahārgāh}, \emph{māhūr}, and \emph{rāstpanjgāh}.


The \emph{āvāzes} are: \emph{bayāt-e-kord}, \emph{bayāt-e-tork} (also referred to as \emph{bayāt-e-zand}), \emph{dashtī}, \emph{abū'atā}, \emph{afshārī}, and \emph{bayāt-e-esfahān}. 

Among the six \emph{āvāzes}, \emph{bayāt-e-esfahān} is a subcategory of the \emph{homāyūn}, while the remaining are subcategories of the \emph{shur}. In many accounts, \emph{radif} is considered to have 5 \emph{āvāzes}, as \emph{bayāt-e-kord} is often omitted. The reason is that most experts dispute the requirement of recognizing it as a independent \emph{āvāz}. In this study, we have included \emph{bayāt-e-kord} to ensure a complete representation.

Each of these subcategories comprises pieces called \emph{gūshehs}. These \emph{gūshehs} can range from being as brief as a single sentence to as extensive as a full composition, with performances lasting several minutes. 

\emph{Gūshehs} can be divided into three types: modal, melodic, and rhythmic. Modal \emph{gūshehs} are played to introduce a mode as a small framework for improvisation. Melodic \emph{gūshehs} introduce a specific melody and its variations, where that specific melody remains fixed in different performance versions. Rhythmic \emph{gūshehs} represent a specific rhythm and its variations. The same \emph{gūsheh} names may appear in different \emph{dastgāhs} or \emph{āvāzes}. \emph{Kereshmeh} is a rhythmic \emph{gūsheh} that appears multiple times in the \emph{radif}, sharing the same rhythmic pattern in each case. Another example is \emph{hazīn}, a melody that is performed in different modes; it is classified as a melodic \emph{gūsheh}. \emph{Qaracheh} is an example of a modal \emph{gūsheh} that appears in more than one \emph{dastgāh}.

The term \emph{āvāz} has three meanings: 1) broadly, it refers to singing; 2) more generally, it refers to Iranian non-metric music; and 3) more specifically, it signifies the segments of \emph{radif} that are smaller than a \emph{dastgāh}. This study focuses on the third definition, though the other two meanings are clarified where relevant.

Non-metric music refers to musical organization that lacks regular meter while potentially maintaining other temporal structures \cite{lerdahl1983}. The distinction between non-metric music and free-rhythm music centers on the preservation of proportional durational relationships. \cite{clayton1996} defines free rhythm as ``the rhythm of music without perceived periodic organization,'' encompassing music where temporal organization serves non-rhythmic goals such as text transmission or melodic exposition. We consider that \textbf{non-metric music} maintains relative proportional relationships between note durations despite lacking metrical organization, while \textbf{free-rhythm music} may abandon proportional consistency entirely. 

Tsuge discusses the concept of non-metric music and emphasizes its greater importance in Iranian music compared to other traditions~\cite{tsuge1970rhythmic}. He explains that the rhythmic structure of \emph{āvāz} (second definition) music is mainly based on the poetic rhythm system, where a repeating pattern of different number and size of syllables shapes its rhythm. This structure is closely connected to the nature of the Persian (Farsi) language and its classical poetry system which plays a significant role in how the melody is formed and perceived. Kanani and Azadehfar~\cite{kanani2020study} described the key \emph{āvāz} (second definition) patterns commonly found in non-metric traditional Iranian vocal music.

The exact origins of the \emph{radif} system in Iranian music are not clearly defined. Some sources, like Bruno Nettl, believe it originated in the 17th century, while others suggest the 18th century as the starting point~\cite{nettl2012music,miller1993radif,talai1993new}. What is clear, however, is that \emph{radif} developed from the late Safavid era (1670s-1730s) through to the mid-Qājār period (1850s). The lack of precise dating can be linked to the oral tradition of this music and the absence of recording technology at the time. 

It is believed that \emph{radif} was created to support the teaching of musical modes and to enhance skills in improvisation and modulation within Iranian art music~\cite{farhat2004dastgah}. The same \emph{radif} can be interpreted differently by different musicians, and once a student becomes a master, they are able to develop their own version of the \emph{radif}. Over time, many prominent music masters have created their own interpretations, leading to different versions of \emph{radif}. These musicians developed their \emph{radif} based on their personal understanding, experience, and expression of Iranian modes, either for their own performances or to teach younger learners. Traditionally, \emph{radif} was passed down orally from master to student, preserving its legacy and technical details through generations.

The version of \emph{radif} curated by musician and educator Mīrzā 'Abdollāh has become the most widely used choice in private lessons, university music education and conservatories over the past century. Initially, it was mainly associated with the \emph{tār} and \emph{setār} instruments, but today, it has been adapted and performed on many key Iranian instruments, including \emph{kamancheh}, \emph{santūr}, \emph{ney}, \emph{qānūn}, \emph{oūd}, and \emph{qeychak}.

\subsection{Exploring Datasets}

In recent years, the creation and sharing of digital music corpora has gained significant attention among researchers in areas such as music information retrieval, computational musicology, and natural language processing. Various studies have demonstrated that well-curated datasets can facilitate analysis of both symbolic and audio musical features, thereby promoting new insights into musical traditions~\cite{papaioannou2022dataset, de2019josquintab}, styles, and technologies. 

Many music information retrieval (MIR) corpora are primarily audio-based, with annotations for pitch, timing, structural information, etc., e.g.~\cite{foster2021filosax, beauguitte2016corpus}, while others are symbolic / score-based. Of the latter, some are based on automated reading of paper scores~\cite{polykarpidis2022three}. Ours differs in that we have manually written the digital score.

Considering other musical traditions in the geographical region, there is no symbolic corpus available for Arabic Maqām music, whereas a symbolic corpus does exist for Turkish Makams, known as SymbTr~\cite{karaosmanouglu2012turkish}.

There are some audio datasets related to these musical traditions, such as the {\em Dunya} corpus, which includes Turkish Makam~\cite{uyar2014corpus}, Carnatic, Hindustani~\cite{srinivasamurthy2014corpora}, Beijing Opera~\cite{repetto2014creating}, and Arab-Andalusian music~\cite{sordo2014creating}. The Dunya corpus is part of a larger project called CompMusic~\cite{serra2014creating}.


In~\cite{shafiei2021extracting}, the authors introduce a MIDI dataset for the Mīrzā 'Abdollāh \emph{radif}, which, to the best of our knowledge, is the first study of its kind in this field. Furthermore, in our previous work, we introduced the Shour Corpus~\cite{kanani2024grammatical}, which is used in our study on discovering patterns and producing meaningful variations through grammatical representation (compression) in this musical style.

KUG Dastgāhi~\cite{nikzat2022kdc} and~\cite{heydarian2005database} are two audio datasets for Iranian music. Nava~\cite{baba2019nava} is an audio dataset designed for Iranian instrument recognition, while Ar-MGC is a dataset for Arabic music genre classification~\cite{almazaydeh2022arabic}.

\subsection{Our Contribution}

\begin{figure}[h]
  \centering
  {\includegraphics[width=0.47\textwidth]{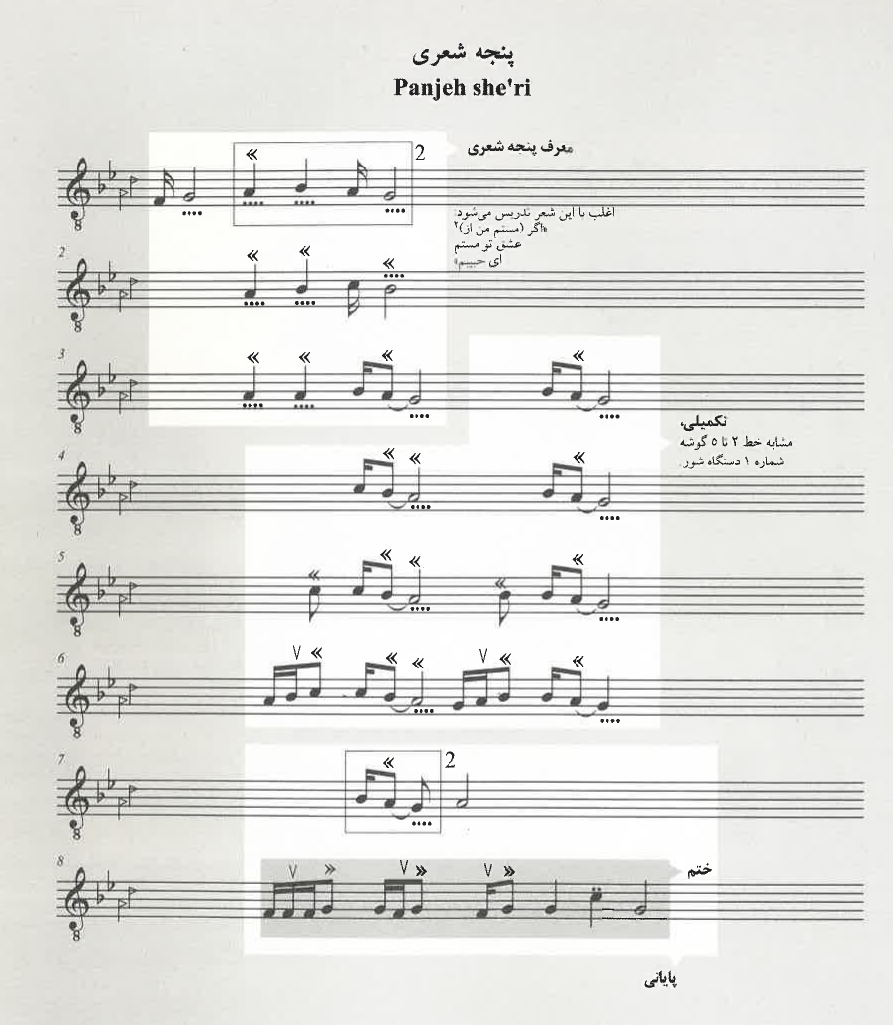}}
  \caption{Transcription sample from Dariush Talai's ``Radif Analysis," illustrating the second \emph{gūsheh} structure in the \emph{shur} \emph{dastgāh}. The three white boxes indicate the main structural divisions, while the last white box contains a further sub-division marked in gray.\label{fig:transcript}}
\end{figure}

A symbolic dataset encompassing the full \emph{radif} along with multiple representations, hierarchical structure, and statistical metadata has been lacking in the literature. This led us to create the Radif Corpus, which includes all non-metric pieces from Mīrzā 'Abdollāh’s \emph{radif}. Out of the several transcriptions of Mīrzā 'Abdollāh's \emph{radif}, we have selected the edition titled ``Radif Analysis - based on the notation of Mīrzā 'Abdollāh's \emph{radif} with annotated visual description'' by Dariush Talai~\cite{talai2015}. This edition consists of a recorded performance, together with a score derived from the performance, notated with hierarchical structure. Although \emph{radif} also includes some metric pieces, which are usually performed at the end of each \emph{dastgāh}/\emph{āvāz}, our corpus excludes them as we are focusing on non-metric music. Figure~\ref{fig:transcript} presents an example of a transcription from the book.

In~\cite{serra2014creating}, Serra identified five critical criteria for corpora in the CompMusic project: Purpose, Coverage, Completeness, Quality, and Reusability. These are criteria that we also considered in our work. Our purpose has already been stated; Coverage and Completeness are achieved by including an entire \emph{radif} collection. Regarding Quality, we believe the transcription is accurate, as it has been double-checked by one of the authors who is an expert in this musical style. Reusability is addressed by providing open data in a documented format.

This corpus is a resource for MIR, computational musicology, and ethnomusicology, enabling applications such as melodic pattern recognition, automatic transcription, and mode classification. It enables symbolic music generation and AI-assisted improvisation. The dataset also facilitates cross-cultural music studies, allowing for comparative analysis with Turkish Makams and Arabic Maqams, as well as phrase-level examinations of melodic progression. Additionally, its structured format aids computational analysis of non-metric rhythm and hierarchical structure understanding, making it a foundational dataset for exploring Iranian classical music in both traditional and computational domains. 

\section{Radif Corpus Description}

Our corpus includes all these \emph{dastgāhs}/\emph{āvāzes}, featuring 228 non-metric \emph{gūshehs}. The MIDI files contain a total of 43,441 notes, with a total playback duration of approximately 16,825 seconds (about 281 minutes).

The dataset represents each musical piece as a sequence of notes, where for each note we store microtonal pitch, duration, pitch (quarter tones), interval, MIDI pitch number and MIDI bend. Data formats are csv files and MIDI files, which have been manually transcribed from the book. 

Additionally, we provide MusicXML files converted from the CSV data. These XML files preserve the microtonal pitch information using fractional \texttt{<alter>} values following MusicXML 4.0 standards. For non-metric rhythm representation, we use flexible time signatures that accommodate the total duration of each piece. However, we note that some current music notation software implementations show limitations in both microtonal playback and non-metric representation. We observed that quarter-tones are not played back correctly, and the software tends to generate complex time signatures (e.g., 342/8) as a workaround for representing non-metric music, which, while functional, may not provide an aesthetically ideal notation display. The MusicXML conversion script is included in the repository for researchers who wish to experiment with different notation software or contribute to improving microtonal MusicXML rendering capabilities. These files don't represent hierarchical structures.

The dataset includes several figures that are explored further in the continuation of this paper. Our digital version exactly mimics the paper source, while to simplify the dataset and avoid additional complexity, grace notes or ornaments are not included in the dataset.

The accurate representation of Iranian classical music involves dealing with two main issues: non-metric rhythm and micro-tonal pitch. In the following subsections, we describe our methods to address these challenges.

\textbf{Tonality.} 
Notes are symbolized by ${C, D, E, F, G, A, B}$, with accidental signs including flat (\musFlat{}), \emph{Koron} ($k$), \emph{Sori} ($s$), and sharp (\musSharp{}). Here, ``Koron'' and ``Sori'' indicate micro-tonal adjustments specific to Iranian music - quarter tones lower and higher, respectively.

\textbf{Chromatic Scale.}
Although these intervals suggest a 24-quarter-tone chromatic scale per octave, which can be seen in some contemporary compositions, Iranian  instruments traditionally employ only 18 specific notes:
{C, D\musFlat{}, D$k$, D, E\musFlat{}, E$k$, E, F, Fs, F\musSharp{}, G$k$, G, A\musFlat{}, A$k$, A, B\musFlat{}, B$k$, B},
with corresponding quarter-tone intervals:
{2, 1, 1, 2, 1, 1, 2, 1, 1, 1, 2, 1, 1, 2, 1, 1}.

\textbf{MIDI.}
Pitch bend is a commonly used method for representing microtones in MIDI files in microtonal music styles. To encode Koron, we assign it a MIDI note number one semitone lower than the natural note and a pitch bend, i.e.~increase of 2048 (a quarter tone); for Sori, the MIDI note number is the same as the natural, with a pitch bend increase of 2048.

\textbf{Octaves.}
We consider the lowest note in the first \emph{gūsheh} of each \emph{dastgāh}/\emph{āvāz} as the start note of the main octave. For example in the first \emph{gūsheh} of our corpus the lowest note is F3 so the main octave is from F3 to F4. The music in the main octave is represented only by its symbols and accidental signs (if any are present). For notes in octaves other than the main one, we use ‘+’ or ‘-’ followed by a number to indicate the number of octaves above or below the main octave. For example, an A$k$ from one octave higher than the main octave would be written as A$k$+1.

\textbf{Intervals.}
The Intervals column represents the pitch difference between consecutive notes, with "1" indicating a quarter-tone step.

\textbf{Durations.}
The term non-metric does not mean the same as free rhythm. In this musical style, notes are related to each other through proportional duration differences, with some being longer or shorter than others. These relationships create the rhythmic structure, which is mostly fixed and not altered by the performer. While the exact durations are not strictly defined, they can be categorized into four main types: very short, short, long, and very long. These can be said to correspond to sixteenth, eighth, quarter, and half notes~\cite{talai2015} and are numerically represented as 1, 2, 4, and 8 in the corpus, where 1 rhythmic unit is equivalent to 1 sixteenth note.

\textbf{Greater Hierarchical Structures.}
We also documented the hierarchical structure of each piece, as provided in the original printed source (see Figure~\ref{fig:transcript}). In our notation, brackets represent hierarchical relationships, forming a tree structure. An open-bracket ``['' in the datasheet marks the beginning of a tree node, with following notes representing the contents of a section or subsection until the matching close-bracket ``]''. Each tune is enclosed within brackets, representing the root node. Additional pairs of brackets define child nodes, which can themselves contain further subsections, forming a nested hierarchy. 

For example, in Figure~\ref{fig:transcript}, the abstract hierarchical structure can be represented as \verb+[ [ ] [ ] [ [ ] ] ]+. The outer brackets enclose the entire tune. The second pair of brackets defines the first section, covering the first three lines in Figure~\ref{fig:transcript}. The third pair corresponds to the section spanning lines three to six. The next open bracket is followed by another open bracket, indicating the presence of a subsection, which corresponds to lines eight and nine. The subsection is highlighted in the last line. 


\section{Statistical and Visual Overview}

\begin{figure}
  \centering
  \includegraphics[width=0.4\textwidth]{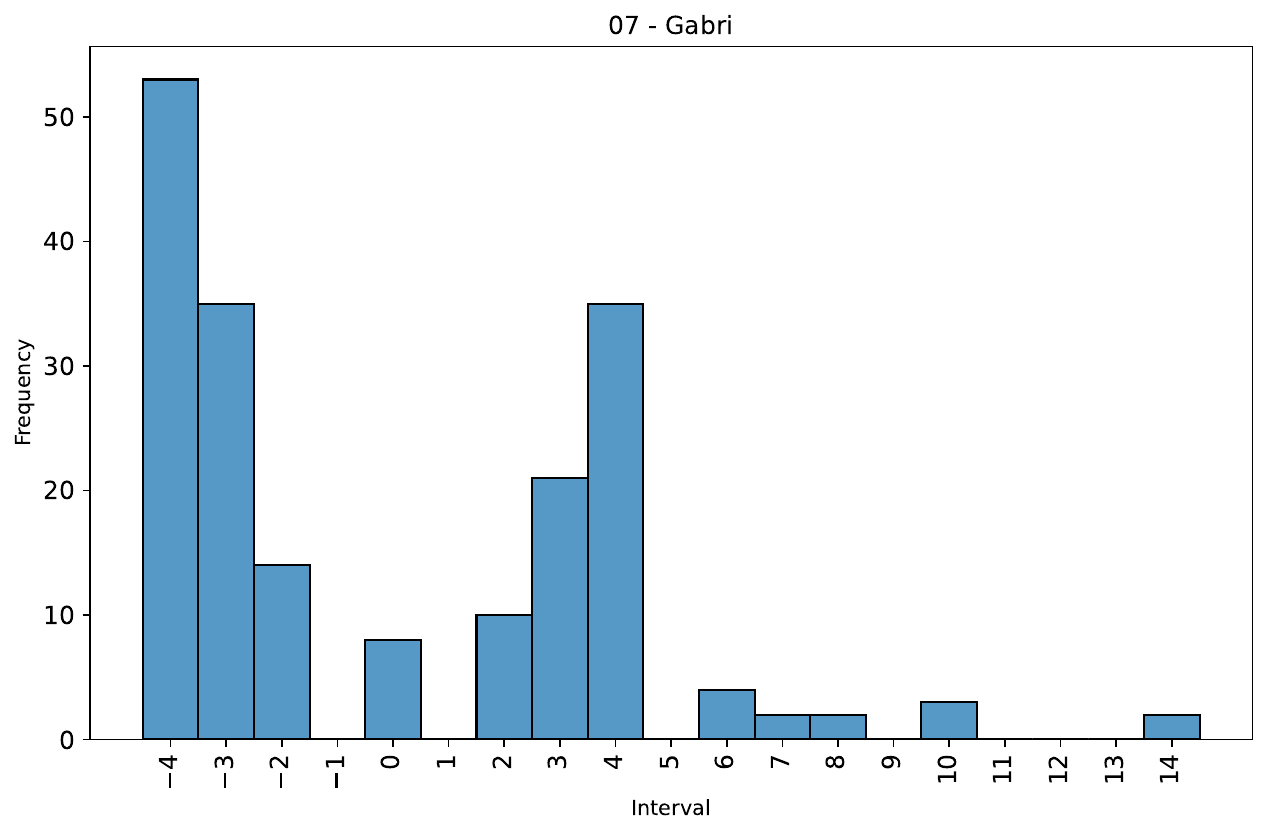}
  \caption{Interval histogram of \emph{Gabri}, a \emph{gūsheh} in \emph{abū'atā}\label{fig:Gabri}}
\end{figure}

In the dataset, for each \emph{gūsheh}, we provide both a pitch histogram and an interval histogram. Figure~\ref{fig:Gabri} provides an example of the interval histogram for a \emph{gūsheh}.

Each \emph{dastgāh} or \emph{āvāz} comes with a spreadsheet giving information about its \emph{gūshehs}, like the number of notes and total duration. Table \ref{tab:primary statistics} presents the number of \emph{gūshehs} in each \emph{dastgāh} or \emph{āvāz}, along with the number of notes, duration in both units and seconds, and pitch range.  

\emph{Māhūr} has the highest number of \emph{gūshehs} with 34, followed by \emph{chahārgāh} with 31 and \emph{shur} with 29. It also contains the largest number of notes, with 6104 in \emph{māhūr}, 5788 in \emph{chahārgāh}, and 4830 in \emph{shur}. These three also have the longest durations, with values of 12480, 11606, and 9839 rhythmic units, respectively.

\emph{Afshārī} has the fewest \emph{gūshehs} with 4, followed by \emph{bayāt-e-esfahān} with 5, and \emph{dashtī} and \emph{bayāt-e-kord} with 6 each. The shortest \emph{dastgāh}/\emph{āvāz} in terms of duration is \emph{dashtī}, with 1308 notes and a duration of 2685 units, followed by \emph{bayāt-e-kord} with 1360 notes and a duration of 2432 units. \emph{afshārī} is the third shortest, with 1520 notes and a duration of 2681 units.  

Regarding individual \emph{gūshehs}, \emph{kereshmeh} in \emph{segāh} is the shortest \emph{gūsheh} in the corpus, while \emph{bayāt-e rāje' va forūd} in \emph{bayāt-e-esfahān} is the longest.

\begin{table}[htbp]
  \centering
  \resizebox{\columnwidth}{!}{%
    \begin{tabular}{|p{2.4cm}|p{1cm}|p{1cm}|p{1.4cm}|p{2cm}|p{1.8cm}|}
      \hline
      \emph{dastgāh}/\emph{āvāz} & \emph{gūsheh} Count & Number of Notes & Total Duration (unit) & MIDI Performance Duration (second) & Pitch Range\\
      \hline
      \emph{Shur} & 29 & 4830 & 9839 & 1966 & [F, A\musFlat{}+2] \\
      \hline
      \emph{Bayāt-e-kord} & 6 & 1360 & 2432 & 486 & [G-1, A\musFlat{}+1] \\
      \hline
      \emph{Dashtī} & 6 & 1308 & 2685 & 536 & [F-2, G+1] \\
      \hline
      \emph{Bayāt-e-tork} & 16 & 2544 & 4986 & 996 & [F-1, G+1] \\
      \hline
      Abuata & 7 & 2194 & 3959 & 791 & [F, A\musFlat{}+1] \\
      \hline
      \emph{Afshārī} & 4 & 1520 & 2681 & 536 & [F-1, A\musFlat{}+1] \\
      \hline
      \emph{Segāh} & 20 & 3283 & 6559 & 1310 & [F, F+2] \\
      \hline
      \emph{Navā} & 19 & 3252 & 5975 & 1194 & [D-1, C+1] \\
      \hline
      \emph{Homāyūn} & 27 & 5323 & 9780 & 1954 & [D, F+2] \\
      \hline
      \emph{Bayāt-e-esfahān} & 5 & 1669 & 3264 & 652 & [D-1, F\musSharp{}+1] \\
      \hline
      \emph{Chahārgāh} & 31 & 5788 & 11606 & 2319 & [C-1, G+2] \\
      \hline
      \emph{Māhūr} & 34 & 6104 & 12480 & 2493 & [C-1, G+2]\\
      \hline
      \emph{Rāstpanjgāh} & 24 & 4266 & 7958 & 1590 & [D-1, C+2]\\
      \hline
      
    \end{tabular}%
  }
  \caption{Summary of \emph{gūsheh} information for each \emph{dastgāh} and \emph{āvāz}, including the number of \emph{gūshehs}, total notes, duration in units and seconds and pitch range.}
  \label{tab:primary statistics}
\end{table}

\subsection{Melodic Progression}

One of the main objectives when a musician performs a complete concatenated \emph{dastgāh} is to follow \emph{seyr}, or melodic movement \cite{chalesh2017seyir}. In traditional Iranian music, \emph{seyr} refers to the progression of melodies within a piece, shaping the overall pitch direction of the music through its introduction, development, climax, and resolution.

\emph{Seyr} underlines the importance of transitional notes and melodic phrases in establishing the identity and modal character of the piece. These elements play a crucial role in guiding the melodic flow from one section to another, ensuring a coherent and expressive musical journey.

Each \emph{dastgāh}/\emph{āvāz} folder includes a pitch contour plot to show its \emph{seyr}. The pitch contour plot illustrates how the melody and pitch evolve across different \emph{gūshehs}.


\subsection{Analyzing the Functionality of Pitches}

One of the other key differences between Iranian classical music and Western music lies in how the central pitch is treated. In Western music, the tonic (or root note) is the main pitch around which a key is built, providing a sense of stability and resolution. In contrast, Iranian music uses the concept of \emph{shāhed}, a pitch that is emphasized or serves as a focal point within an \emph{āvāz} or \emph{dastgāh}, but does not necessarily act as the final resting note. The \emph{shāhed} can shift or be re-emphasized within a \emph{dastgāh}/\emph{āvāz} performance. Additionally, Persian music distinguishes between the \emph{shāhed} (point of focus), \emph{Ist} (stops or longer rests through phrases of melody) and \emph{Khatemeh} (cadential notes), while in Western theory, the tonic usually plays all  three roles. 
Due to these differences, the \emph{shāhed} and tonic may occasionally coincide, but they represent distinct concepts. Understanding this distinction is essential for analyzing the structure and phrasing of Iranian classical music. We don't annotate any of these concepts in the dataset as there is some ambiguity in them and they are not provided in the source.

\begin{figure*}
  \centering
  \includegraphics[width=0.6\textwidth]{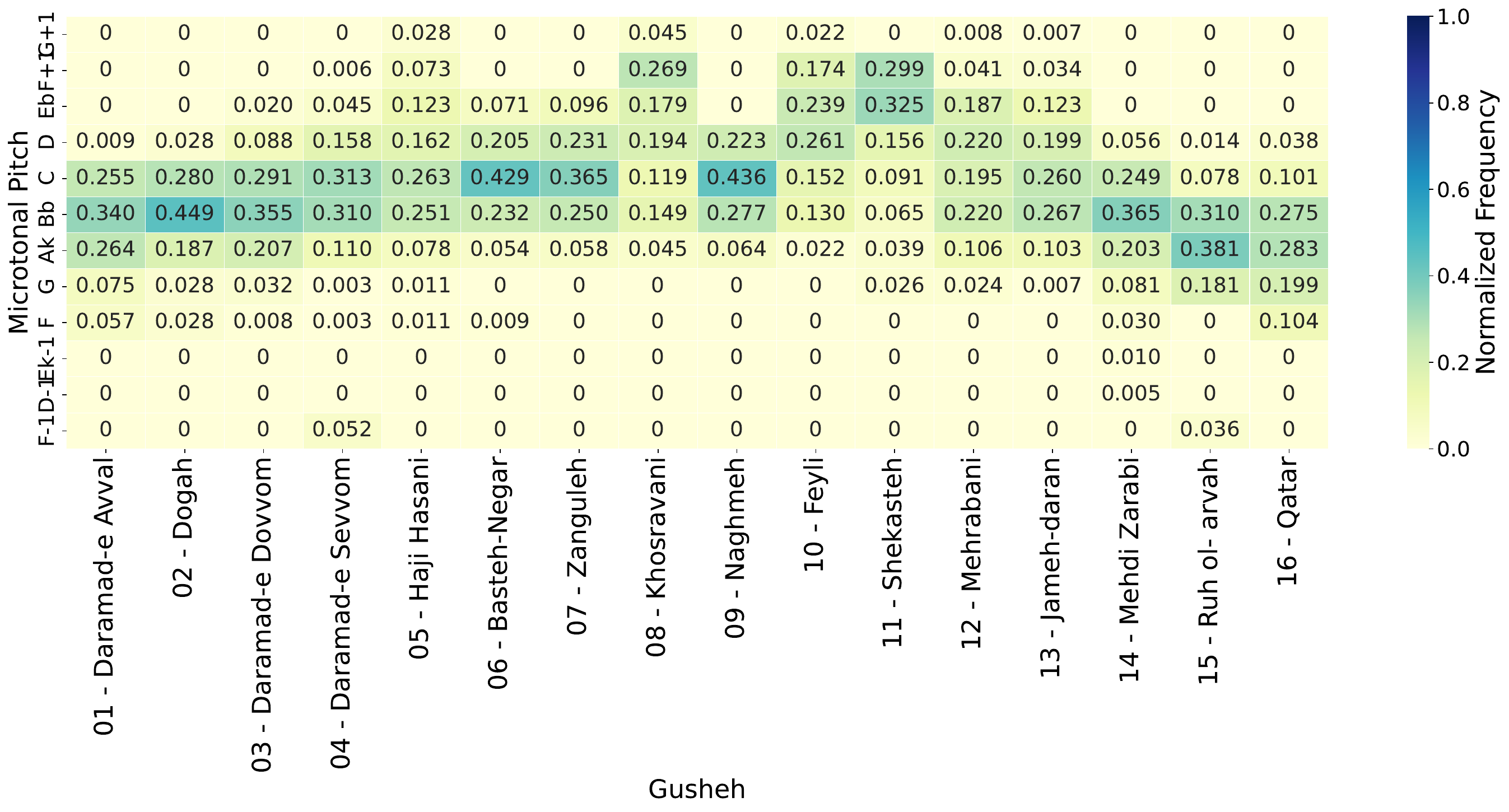}
  \caption{Heat map of pitch occurrences across 16 \emph{gūshehs} in \emph{bayāt-e-tork} \emph{āvāz}, shown along the time axis, illustrating tonal functionality in each \emph{gūsheh} and shifts in the most repeated tone throughout the melodic progression.\label{fig:Tork}}
\end{figure*}

Fig.~\ref{fig:Tork} shows the frequency of occurrence of each pitch in each tune. This plot can be adapted for studying pitch functionality in each \emph{gūsheh}. It also shows how the most repeated pitch has shifted. This heat map can be found in the dataset for each \emph{dastgāh}/\emph{āvāz}.


\subsection{Complexity Analysis}

\begin{figure}
  \centering
  \includegraphics[width=0.48\textwidth]{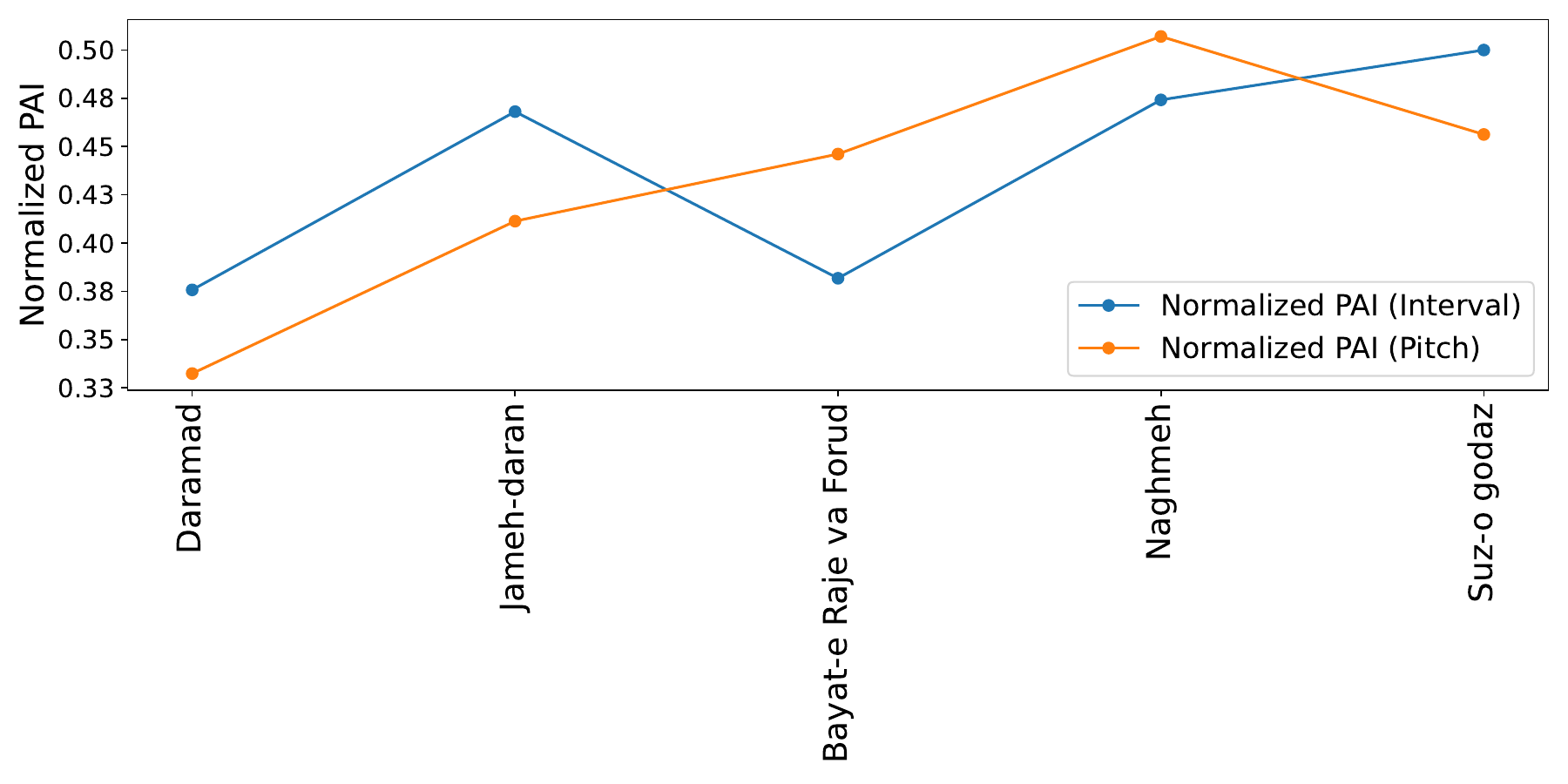}
  \caption{Normalised complexity across \emph{bayāt-e-esfahān}\label{fig:PAI}}
\end{figure}

McCormack et al.~\cite{mccormack2021enigma} argue that complexity is fundamental to human creativity and observe that most art occupies a middle ground—neither too simple nor too complex. Our previous works~\cite{kanani2023graph, kanani2023parsing, kanani2024grammatical} and ongoing research attempt to compare musical complexity across different pieces. We calculated the normalized Pathway Assembly Index (PAI)~\cite{marshall2019quantifying} to assess the complexity of each \emph{gūsheh}. The PAI represents the number of steps needed to reconstruct a melody through the binary concatenation of pitches and previous concatenations (e.g., \texttt{\{a, b, c, d, r\} -> ca -> ab -> ra -> cad -> abra -> cadabra -> abracadabra}~\cite{marshall2019quantifying} gives PAI=7). The normalized PAI is the PAI divided by the tune length. Figure~\ref{fig:PAI} displays the normalized PAI for \emph{āvāz} \emph{bayāt-e-esfahān}, where we calculate PAI over the tune represented as pitches, and then separately over the tune represented as intervals. Across the dataset, the average PAI is 0.47. Specifically, \emph{shur} (pitch) shows the highest complexity at 0.51, while \emph{afshārī} (pitch) records the lowest at 0.36. Higher complexity means less repetitive patterns. We didn't observe any consistent trend in complexity within \emph{dastgāh}/\emph{āvāz}.

\subsection{Similarity Analysis}

An interesting aspect is the comparison of similarities within each \emph{dastgāh}/\emph{āvāz} and across the entire dataset. We analyzed similarity using normalized Damerau-Levenshtein distance, a method commonly used for melody similarity comparison~\cite{fonn,polifoniaD3.2}. Specifically, pitch sequences were compared within each \emph{dastgāh}/\emph{āvāz}, and interval sequences were used for comparisons across the entire dataset. This approach helps to limit the effects of musical transposition and identifies similar motifs' movements that appear in different sections of the \emph{radif}. 

\urldef\similaritylink\url{https://limewire.com/d/IXH5G#ZUZyGVbiqi}

\begin{figure*}[h!]
  \centering
  \includegraphics[trim=800 0 400 0,clip,width=0.65\textwidth]{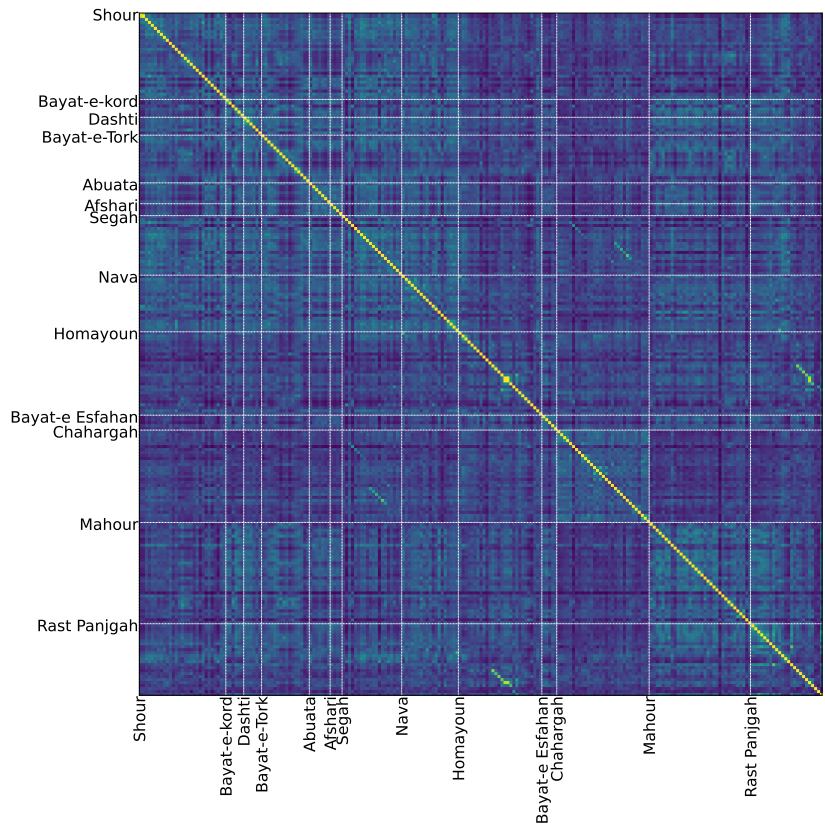}
  \caption{Similarity matrix across the corpus. Light colours indicate high similarity.}
  \label{fig:similarities}
\end{figure*}

Figure~\ref{fig:similarities} presents a similarity matrix for the 228 \emph{gūshehs} in the dataset, with corresponding matrices for each individual \emph{dastgāh}/\emph{āvāz} available in the corpus.\footnote{High-resolution version of the similarity matrix: \similaritylink}

We added horizontal and vertical lines to visually separate each \emph{dastgāh}/\emph{āvāz}. This figure highlights notable musical resemblances: the short diagonal lines e.g.~near bottom-centre indicate strong similarities among some \emph{gūshehs} in \emph{chahārgāh} and \emph{segāh}, as well as among some in \emph{homāyūn} and \emph{rāstpanjgāh}. 

The large square block in the top-left corner suggests that all \emph{dastgāhs}/\emph{āvāzes} within that region (from \emph{shūr} to \emph{navā}) share significant similarities. Furthermore, the figure effectively illustrates the self-similarity within most \emph{dastgāh}/\emph{āvāz}. All except \emph{homāyūn} form distinct blocks along the diagonal line that represent these internal similarities. The block in the bottom-right confirms the similarity between \emph{māhūr} and \emph{rāstpanjgāh}.

A prominent point along this diagonal corresponds to several \emph{gūshehs} with strong mutual resemblance. In \emph{homāyūn}, several \emph{gūshehs} are named after \emph{norūz} (\emph{norūz ‘arab’}, \emph{norūz sabā}, and \emph{norūz khārā}), and parts of these are highly similar. 
These findings could pave the way for further studies based on these musical similarities matrices in future research.

\section{Availability and Access}
The Radif Corpus is openly available on Zenodo under the DOI: 10.5281/zenodo.15742125. It includes the musical data in CSV MIDI, and MusicXML formats. In addition, it contains supporting figures, pitch and interval histograms, pitch contour plots, and similarity matrices for each\emph{dastgāh} and \emph{āvāz}.

All materials are released under a CC-BY 4.0 license, allowing for reuse and adaptation with proper credit. Researchers are encouraged to cite this paper when using the dataset in academic or creative work.

\section{Conclusion}

In this paper, we have introduced the Radif Corpus, a comprehensive symbolic dataset for Iranian classical music that covers all 13 \emph{dastgāh}/\emph{āvāz} of this tradition, specifically the non-metric pieces from Mīrzā 'Abdollāh's \emph{radif}. This dataset addresses the non-metric core of Iranian art music by documenting microtonal pitch, melodic progressions, and hierarchical structures across 228 pieces. By offering detailed annotations in both MIDI and CSV formats, the Radif Corpus provides a valuable platform for research in music information retrieval, music theory, and computational (ethno)musicology. We hope it will inspire future explorations of Iranian classical music and serve as a foundational resource for advancing analytical and creative studies in this domain.

\section{Acknowledgments}

This work was conducted with the financial support of the Research Ireland Centre for Research Training in Digitally-Enhanced Reality (d-real) under Grant No. 18/CRT/6224. For the purpose of Open Access, the author has applied a CC-BY public copyright license to any Author Accepted Manuscript version arising from this submission.

\section{Ethics Statement}

Our work uses publicly available music corpora and does not involve human participants directly. The music collection is part of the Iranian classical music tradition, which is not itself owned or copyrighted by any individual composer. The edition we have used is used by permission of the author.

This musical tradition is culturally important, and we have tried to be respectful and careful in how we talk about it and use it.

The authors declare no conflicts of interest.

\bibliographystyle{IEEEtran}
\balance
\bibliography{ISMIRtemplate}

\end{document}